\documentclass{appolb}
\usepackage{epsfig}
\usepackage[intlimits]{amsmath}
\usepackage{amsfonts}
\usepackage{amssymb,amscd}

\newcommand{\beq}{\begin{equation}}
\newcommand{\eeq}{\end{equation}}
\newcommand{\beqa}{\begin{eqnarray}}
\newcommand{\eeqa}{\end{eqnarray}}
\newcommand{\be}{\begin{equation}}
\newcommand{\ee}{\end{equation}}

\newcommand{\vp}{\vec{p}}
\newcommand{\vq}{\vec{q}}
\newcommand{\vk}{\vec{k}}
\newcommand{\op}{\omega_{p}}
\newcommand{\oq}{\omega_{q}}
\newcommand{\ok}{\omega_{k}}

\newcommand{\intk}{\sum_{n_k} \int \frac{d^3k}{(2\pi)^3}}

\begin{document}
\title{Dual order parameters and the deconfinement transition
\thanks{Presented by CF at the EMMI Workshop and XXVI Max Born Symposium - Three Days of Strong Interactions,
09th-11th of July 2009 in Wroclaw, Poland}
}
\author{Christian S. Fischer
\address{Institute for Nuclear Physics, 
 Technische Universit\"at Darmstadt,\\ 
 Schlossgartenstra{\ss}e 9, 64289 Darmstadt, Germany\\
 GSI Helmholtzzentrum f\"ur Schwerionenforschung GmbH,\\ 
 Planckstra{\ss}e 1,  64291 Darmstadt, Germany.\\
        E-mail: christian.fischer@physik.tu-darmstadt.de}
\and
Jens~A.~Mueller
\address{ Institute for Nuclear Physics, 
 Technische Universit\"at Darmstadt,\\ 
 Schlossgartenstra{\ss}e 9, 64289 Darmstadt, Germany\\
        E-mail: jens.mueller@physik.tu-darmstadt.de}
}
\maketitle
\begin{abstract}

We investigate the chiral and the deconfinement transition within the framework of 
Dyson-Schwinger equations using quenched lattice data for the temperature dependent
gluon propagator as input. We extract corresponding order parameters from the Landau 
gauge quark propagator with $U(1)$-valued boundary conditions. We study the chiral 
transition using the conventional quark condensate, whereas for the deconfinement 
transition we determine the dual condensate ('dressed Polyakov loop'). In addition 
we consider an alternative order parameter for deconfinement, the dual scalar quark 
dressing function. As a result we find almost the same transition 
temperatures for the chiral and deconfinement transitions.

\end{abstract}
  
\section{Introduction}
The phases of QCD are currently under intense theoretical and experimental investigations.
Open questions concern among others the interplay between the confinement/deconfinement 
transition and the chiral transition, e.g. the (non-)coincidence of the
chiral and the deconfinement transition at zero chemical
potential \cite{Bazavov:2009zn}, and the possibility of a confined chirally
symmetric ('quarkyonic') phase \cite{McLerran:2007qj}. Answers to these
questions certainly require non-perturbative approaches to QCD.

Functional methods involving the 
renormalization group equations \cite{Berges:2000ew} and/or
Dyson-Schwinger equations (DSE) \cite{Roberts:2000aa,Fischer:2006ub} constitute
a non-perturbative continuum approach to QCD.
 Only very recently, approaches accounting also for the deconfinement transition
 became available for these methods \cite{Marhauser:2008fz,heid,Fischer:2009wc,fimu}.
These approaches allow to define order parameters for
the deconfinement transition from the quark propagator for generalized boundary 
conditions. Originally introduced within the 
lattice framework \cite{Gattringer:2006ci,Bilgici:2008qy}
it was adapted to functional methods in \cite{heid,Fischer:2009wc,fimu}. 
The quantities calculated in this work signaling the deconfinement transition
 are the dual quark condensate (or 'dressed Polyakov loop') and the dual scalar 
 quark dressing function.

In the following we first recall the defining equations for the 
ordinary and the dual quark condensate and scalar dressing function, then summarize the truncation
scheme used in our DSE calculations before we discuss our results
for the chiral and deconfinement phase transition.

\section{Order parameters for deconfinement}
Consider the full quark propagator at finite temperature given in terms of its Dirac structure by
\beq
 S(\vp,\op) = \left[i \gamma_4\, \op C(\vp,\op) 
 + i \gamma_i \, p_i A(\vp,\op) + B(\vp,\op)\right]^{-1} \,,
\eeq 
with vector dressing functions $A$ and $C$ and scalar dressing function $B$.
The physical, antiperiodic boundary conditions in temporal direction lead to
$\op(n_t) = (2\pi T)(n_t + 1/2)$ for Matsubara frequencies. Generalizing to temporal $U(1)$-valued
boundary conditions $\psi(\vec{x},1/T) = e^{i \varphi} \psi(\vec{x},0)$ results in Matsubara frequencies $\op(n_t) = (2\pi T)(n_t + \varphi/2)$
with boundary angle $\varphi=[0,2\pi[$.

From the $\varphi$-dependent propagator the dual quark condensate introduced in lattice gauge theory Ref.~\cite{Bilgici:2008qy}
\beq \label{cond}
\Sigma_1 = \int_0^{2\pi} \, \frac{d \varphi}{2\pi} \, e^{-i\varphi}\, 
\langle \overline{\psi} \psi \rangle_\varphi
\eeq
can be calculated. 
Due to its close connection to the Polyakov loop it is also called 'dressed Polyakov loop',
see Refs.~\cite{Bilgici:2008qy,Synatschke:2008yt} for more details.

The dual scalar quark dressing function introduced in Ref.~\cite{fimu} is the phase-Fourier-transform of
the $\varphi$-dependent scalar quark dressing function evaluated at lowest Matsubara
frequency and zero momentum
\beq \label{dual_B}
\Sigma_B = \int_0^{2\pi} \, \frac{d \varphi}{2\pi} \, e^{-i\varphi}\, 
B(0,\op(0,\varphi))\,.
\eeq
Both quantities the dual quark condensate and the dual scalar quark dressing function transform under
center symmetry identically as the conventional Polyakov loop and are therefore order parameters for
deconfinement. 

An advantage of the $\varphi$-dependent quark condensate $\langle \overline\psi \psi \rangle_\varphi$ is its direct connection to the ordinary
Polyakov loop in the limit of static quarks \cite{Bilgici:2008qy}.
For vanishing bare quark mass it is well-behaved in the continuum limit whereas at finite quark mass it is quadratically divergent and needs to be properly regularized. Though apparently not linked to the Polyakov loop
 the $\varphi$-dependent scalar quark dressing function $B(0,\op(0,\varphi))$ has the advantage of being a well defined quantity also in the continuum limit.

\section{The Dyson-Schwinger equation for the quark propagator at finite
temperature}

\begin{figure}[t]
\centerline{\includegraphics[width=0.7\columnwidth]{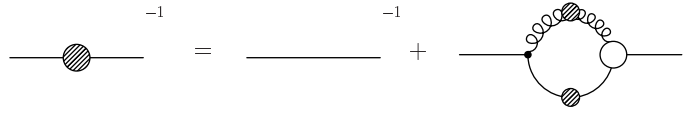}}
\caption{The Dyson-Schwinger equation for the quark propagator. Filled circles
denote dressed propagators whereas the empty circle stands for the dressed
quark-gluon vertex.
\label{fig:quarkDSE}}
\end{figure}
The Dyson-Schwinger equation for the quark propagator 
is displayed diagrammatically in Fig.~\ref{fig:quarkDSE}. At
finite temperature $T$ it is given by 
\beqa \label{DSE}
S^{-1}(p) = Z_2 \, S^{-1}_0(p) 
-  C_F\, Z_{1f}\, g^2 T \intk \gamma_{\mu}\, S(k) \,\Gamma_\nu(k,p) \,
D_{\mu \nu}(p-k) \,, \label{quark_t}
\eeqa
with $p=(\vp,\op)$ and $k=(\vk,\ok)$ and renormalization factors
$Z_2$ and $Z_{1f}$. Here $D_{\mu \nu}$ denotes the (transverse) gluon 
propagator in Landau gauge and $\Gamma_\nu$ the quark-gluon vertex. 
The bare quark propagator is given by $S^{-1}_0(p) = i \gamma \cdot p + m$. 
The Casimir factor $C_F = (N_c^2-1)/N_c$ stems from the color trace; here 
we only consider the gauge group $SU(2)$. The quark dressing functions 
$A,B,C$ can be extracted from Eq.~(\ref{DSE}) 
by suitable projections in Dirac-space. 

In order to solve this equation we have to specify explicit expressions
for the gluon propagator and the quark-gluon vertex. At finite temperatures
the tensor structure of the gluon propagator contains two parts, one 
transversal and one longitudinal to the heat bath. The propagator is then 
given by ($q=(\vq,\oq)$)
\begin{eqnarray}
D_{\mu\nu}(q) = \frac{Z_T(q)}{q^2} P_{\mu \nu}^T(q) 
                    +\frac{Z_L(q)}{q^2} P_{\mu \nu}^L(q) 
\end{eqnarray} 
with transverse and longitudinal projectors 
\beqa
P_{\mu\nu}^T(q) = 
   \left(\delta_{i j}-\frac{q_i q_j}{\vq^2}\right) 
   \delta_{i\mu}\delta_{j\nu}\,, \hspace*{2cm}
P_{\mu\nu}^L(q) = P_{\mu \nu}(q) - P_{\mu \nu}^T(q) \,,
\eeqa 
with ($i,j=1 \dots 3$).
At zero temperatures Euclidean $O(4)$-invariance requires both dressing functions 
to agree, i.e. $Z_T(q)=Z_L(q)=Z(q)$.

\begin{figure}[t]
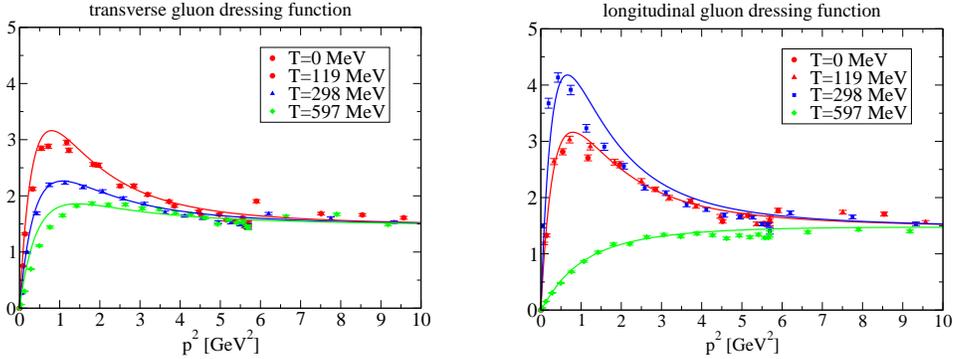

\includegraphics[width=0.45\columnwidth]{glue-trans.eps}\hfill
\includegraphics[width=0.45\columnwidth]{glue-long.eps}
\caption{Quenched $SU(2)$ lattice results \cite{Cucchieri:2007ta} for the 
transverse dressing function $Z_T(q)$ and the longitudinal dressing
function $Z_L(q)$ of the gluon propagator together with the fit 
functions \cite{Fischer:2009wc}.}
\label{fig:lattglue}
\end{figure}

The temperature dependence of the gluon propagator can be inferred from
recent lattice calculations. The results of Ref.~\cite{Cucchieri:2007ta} 
are shown in Fig.~\ref{fig:lattglue}.
The lattice data although still with quantitative uncertanties
 may very well correctly represent the qualitative 
temperature dependence of the gluon propagator. We therefore use a 
temperature dependent (qualitative) fit to the data as input into the DSE; 
this fit is also displayed in Fig.~\ref{fig:lattglue} (straight lines). 
The fit functions are described in detail in Refs.~\cite{Fischer:2009wc,fimu}
and shall not be repeated here for brevity. Note, however, that we also 
inherit the scale determined on the lattice using the string tension 
$\sqrt{\sigma}=0.44$ GeV \cite{Cucchieri:2007ta}.

For the quark-gluon vertex we employ a temperature dependent model which 
is discussed in detail in \cite{fimu}. 
 
\section{Numerical results}
\begin{figure}[t]
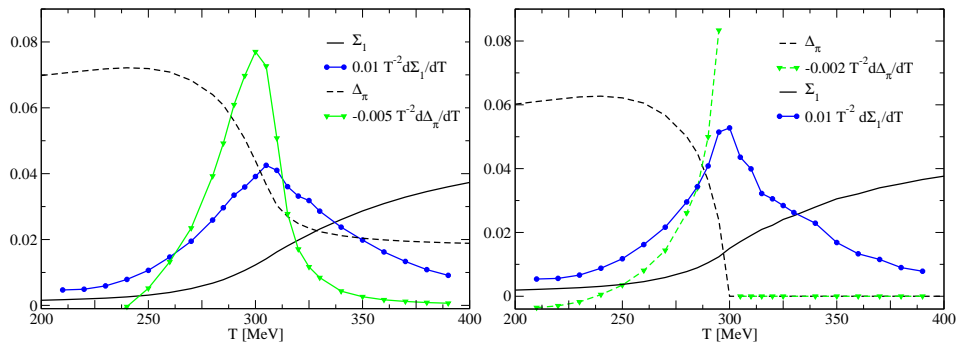

\includegraphics[width=0.5\columnwidth]{cond_dual_cond_vs_T_m10MeV.eps}\hfill
\includegraphics[width=0.5\columnwidth]{Kond_Dual-Kond_vs_Temp_chiral.eps}
\caption{
Left diagram: Temperature dependence of the dressed Polyakov loop 
$\Sigma_1$ and the conventional quark condensate 
$\Delta_\pi \equiv \langle \overline{\psi} \psi \rangle_{\varphi=\pi}$  
together with their derivatives for $m = 10 \,\mbox{MeV}$.
Right diagram: The same quantities in the chiral limit.}
\label{res1}
\end{figure}

In Fig.~\ref{res1} we display our numerical results for the
ordinary and the dual quark condensate together with their (normalized) 
temperature derivatives once evaluated for a quark mass of 
$m = 10 \,\mbox{MeV}$ (fixed at $T=200\,\text{MeV}$ and $\vec{\mu}^2=20\,\text{GeV}^2$) and once evaluated in the chiral limit. One clearly
sees the difference in the chiral transition: whereas at finite bare quark
mass we encounter a crossover the transition changes into a second order
phase transition in the chiral limit. In the first case the corresponding 
temperature derivative shows a peak at $T_c = 301(2)$ MeV, whereas it
diverges at $T_c = 298(1)$ MeV in the second case. We also extracted the
corresponding transition temperatures from the chiral susceptibility
\beq
 \chi_R=m^2\frac{\partial}{\partial m}
 \Big(\langle\bar{\psi}\psi\rangle_T-\langle\bar{\psi}\psi\rangle_{T=0}\Big)\,. 
 \label{def2}
\eeq
The results for quark mass $m = 10 \,\mbox{MeV}$ are given in table \ref{crit_temp}.
\begin{table}[b]
\begin{center}
\begin{tabular}{c|c|c|c}
 $T_c$  & $T_{\chi_R/T^4}$ & $T_{\chi_R}$ & $T_{dec}$   \\\hline\hline
$301(2)$ & $304(1)$         & $305(1)$     & $308(2)$
\end{tabular}
\caption{Transition temperatures for the chiral and deconfinement transition
for quark mass $m = 10 \,\mbox{MeV}$.} \label{crit_temp}
\end{center}
\end{table}

The corresponding transition temperature for the deconfinement transition
can be read off the dual quark condensate (or dressed Polyakov loop). At finite
quark mass and in the chiral limit we observe a distinct rise in the dual 
condensate around $T \approx 300$ MeV. The corresponding (normalized)
temperature derivative shows a peak at $T_{dec} = 308(2)$ MeV for 
quark mass $m = 10 \,\mbox{MeV}$. In the chiral limit this peak moves
to $T_{dec} =299(3)$ MeV. 

In general we note that the chiral and 
deconfinement transition are close together. There are a few MeV difference 
between the different transition temperatures for the crossover at finite 
quark masses, whereas both transitions occur at the same temperature (within errors)
in the chiral limit. These findings agree with early expectations from lattice
simulations \cite{Karsch:1998ua}.

Furthermore we wish to emphasize that the present calculation, although carried
out with quenched lattice results for the gluon propagator, is in itself
not strictly quenched. This can be seen from the fact that
the dressed Polyakov loop is not strictly zero below the deconfinement
transition. We refer here to quenched calculation since the influence of 
the matter content to the gauge sector is neglected.


\begin{figure*}[t]
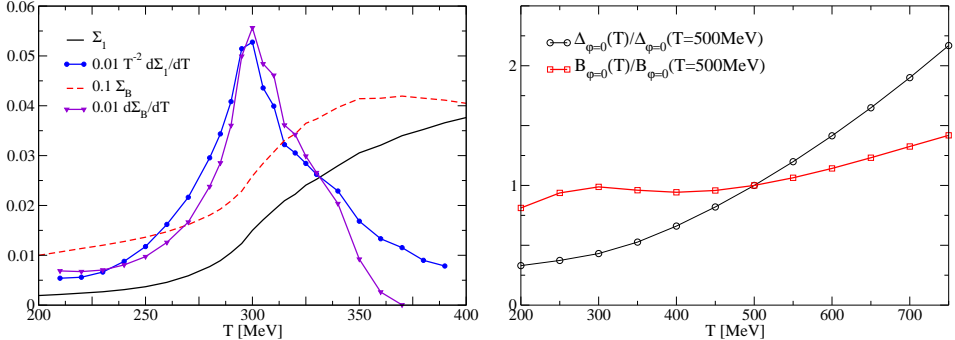

\includegraphics[width=0.5\columnwidth]{dual_cond_dual_b_vs_T_chiral.eps}\hfill
\includegraphics[width=0.47\columnwidth]{Rand_Kond_B.eps}
\caption{
Left diagram: Comparison of the temperature dependence of the 
corresponding dual scalar quark dressing and the dual condensate for $m=10\,\text{MeV}$.
Right diagram: Temperature dependence of the chiral 
quark condensate $\Delta_{\varphi}\equiv\langle\bar{\psi}\psi \rangle_{\varphi}$ and the scalar dressing function
$B_{\varphi}$ at periodic
boundary conditions $\varphi=0$.
}
\label{res:scalar}
\end{figure*}


In the left diagram of Fig.~\ref{res:scalar} we show a comparison of the dual scalar dressing 
Eq.~(\ref{dual_B}) with the dual condensate for $m=10\,\text{MeV}$. Both quantities show a very similar temperature
dependence especially the aforementioned distinct rise around $T \approx 300$ MeV.
The signal is slightly stronger for the dual scalar dressing $\Sigma_B$.
The temperature derivatives indicating the deconfinement transition clearly
 peak at the same transition temperature.

In the plot of the right hand side of Fig.~\ref{res:scalar}
the chiral condensate $\langle\bar{\psi}\psi \rangle_{\varphi=0}$ and
the quark scalar dressing $B(\varphi=0)$
as functions of temperature at periodic boundary conditions are shown.
For comparison both are normalized to $1$ at $T=500$ MeV.
Whereas $\langle\bar{\psi}\psi \rangle_{\varphi=0}$ is a strictly monotonic function with
temperature the scalar dressing is slightly decreasing in the temperature 
range between $300$ and $400$ MeV before the high temperature 
behavior shows up. The large temperature scaling of both quantities
can be extracted analytically from Eqs.~(\ref{cond}) and (\ref{DSE}) 
 as pointed out in the appendix of Ref.~\cite{fimu}. There it is shown that
\beq
B_{\varphi=0, p=0}(T) \sim \sqrt{T} \hspace{1cm}\text{for} \hspace{2mm}T\gg T_c
\eeq
and consequently a quadratic rise of the condensate,
\beqa
 \langle\bar{\psi}\psi \rangle_{\varphi=0} 
  &\sim& T^2\,, \hspace{1cm}\text{for} \hspace{2mm}T\gg T_c\,,
\eeqa
is obtained. These results are in excellent agreement with polynomial fits to the data.
The dual condensate $\Sigma_1$ as well as the dual scalar dressing $\Sigma_B$
are promising candidates for a further study of the deconfinement transition of QCD.

\section{Summary}

In this talk we addressed the chiral and the deconfinement transition 
of quenched QCD. We showed results for the order parameter of the 
chiral transition, the quark condensate, and for order parameters of 
the deconfinement transition, the dressed Polyakov loop and the dual scalar
dressing. These were extracted from the Landau gauge quark propagator evaluated at a continuous
range of boundary conditions for the quark fields. Independent of the dual order parameter used
 we found exactly the same deconfinement transition temperature.
A comparison of the transition temperatures for the chiral and the deconfinement
transition shows almost coincidence for moderate quark mass of the order of an up-quark.
 In the chiral limit the two transitions coincide within error.
We find a second order chiral phase transition at $T_{\chi_R/T^4} = 298(1)$ 
MeV and a similar temperature for the deconfinement transition, 
$T_{dec} =299(3)$ MeV. In summary we conclude that the chiral and deconfinement transition 
temperatures are only slightly different for finite quark masses and coincide within errors
 in the chiral limit.

The framework used in this work is quenched $SU(2)$ Yang-Mills theory.
Our transition temperature may be translated into the corresponding ones
of quenched $SU(3)$ QCD using the relations $T_c/\sqrt{\sigma}=0.709$
($SU(2)$) and $T_c/\sqrt{\sigma}=0.646$ ($SU(3)$) between the respective 
critical temperatures and the string tension \cite{Fingberg:1992ju}. 
The resulting transition temperature is then 
$T_{\chi_R/T^4} \approx T_{dec} \approx 272$ MeV in the chiral limit.
In order to work in the full, unquenched theory quark loop effects and 
meson effects which shift the transition temperature below $T=200$ MeV 
have to be taken into account. Concerning the dual 
condensate and scalar dressing function in the unquenched formulation 
additional effects due to the Roberge-Weiss symmetry \cite{Roberge:1986mm} 
occur. This is because of the formal 
similarity of the continuous boundary conditions for the quark field to an 
imaginary chemical potential, see \cite{Marhauser:2008fz} for details. 

{\bf Acknowledgments}\\
It is a pleasure to thank the organisers of this exciting and inspiring
workshop for their efforts. This work has been supported by the Helmholtz 
University Young Investigator Grant VH-NG-332 and by the Helmholtz Alliance HA216-TUD/EMMI.

\end{document}